\documentclass{aa}
\usepackage{graphicx}
\usepackage[T1]{fontenc}
\usepackage{ae,aecompl}
\usepackage{txfonts}
\begin{document}
   \title{Large amplitude oscillation of a polar crown filament 
   in the pre-eruption phase}

   \subtitle{}

   \author{Hiroaki Isobe \inst{1,2} \and Durgesh Tripathi\inst{1}} 
   \offprints{H.Isobe; e-mail: isobe@eps.s.u-tokyo.ac.jp}

   \institute{Department of Applied Mathematics and Theoretical Physics, 
   University of Cambridge, Wilberforce Road, Cambridge CB3 0WA, UK \\
              \email{isobe: D.H.Isobe@damtp.cam.ac.uk; 
                tripathi: D.Tripathi@damtp.cam.ac.uk}
         \and
           Department of Earth and Planetary Science, University of Tokyo, 
		   Hongo, Bunkyo-ku, Tokyo 113-0033, Japan \\
                   \email{isobe@eps.s.u-tokyo.ac.jp}
       }
   \date{Received 1 February 2006 / accepted 10 February 2006}

   \abstract
{}
{We report observation of a large-amplitude filament oscillation 
followed by an eruption. This is used to probe the pre-eruption condition  
and the trigger mechanism of solar eruptions.}
{We used the EUV images from the Extreme-Ultraviolet Imaging Telescope on board 
SOHO satellite and the H$\alpha$ images from the Flare Monitoring Telescope 
at Hida Observatory. The observed event is a polar crown filament that 
erupted on 15 Oct. 2002.}
{The filament clearly exhibited oscillatory motion in the slow-rising, 
pre-eruption phase. 
The amplitude of the oscillation was larger than 20 km s$^{-1}$, 
and the motion was predominantly horizontal. The period was 
about 2 hours and seemed to increase during the oscillation, indicating 
weakening of restoring force.}
{Even in the slow-rise phase before the eruption, the filament retained 
equilibrium and behaved as an oscillator, and the equilibrium is 
stable to nonlinear perturbation. 
The transition from such nonlinear stability to either  
instabilities or a loss of equilibrium that leads to the eruption occurred 
in the Alfv\'{e}n time scale ($\sim$ 1 hour). This suggests that 
the onset of the eruption was triggered by a fast magnetic reconnection 
that stabilized the pre-eruption magnetic configuration, rather 
than by the slow shearing motion at the photosphere.
}
\titlerunning{Oscillation of erupting filament}
\authorrunning{H. Isobe \& D. Tripathi}
   \maketitle

\newpage

\section{Introduction}
Filament (prominence) eruptions are associated with various kinds of 
solar activity, such as coronal mass ejections 
(CME; Munro et al. \cite{munro}; Webb \& Hundhausen \cite{webb}),  
flares (Hirayama \cite{hirayama}; Kahler et al. \cite{kahler}) , 
giant arcade formations (McAllister et al. \cite{mcallister}; 
Tripathi et al. \cite{tripathia}; Shiota et al. \cite{shiota}), 
and even microflares 
(Sakajiri et al. \cite{sakajiri}). 
Despite the differences in size, energy and morphology, these 
eruptive phenomena in the solar atmosphere may be different 
aspects of a common physical process involving plasma ejection 
and magnetic reconnection (e.g., Shibata \cite{shibata}; 
Priest \& Forbes \cite{priest02}). 

Understanding the triggering and driving mechanisms of the solar 
eruptions is one of the most important issues in current solar 
physics. Numerous theoretical models have been published on this 
subject (see review by Forbes \cite{forbes}).  
In terms of what actually triggers the onset of eruptions, 
some involve 
magnetic reconnection to destabilize the pre-existing magnetic system 
(e.g., Antiochos et al. \cite{antiochos}; Chen \& Shibata \cite{chen}; 
Moore et al. \cite{moore}),  
while others consider either 
shearing/converging motions at the photosphere 
(e.g., Miki\'{c} \& Linker \cite{mikic}; 
Kusano et al. \cite{kusano}; Priest \& Forbes \cite{priest95}; 
T\"{o}r\"{o}k \& Kliem \cite{toeroek}) 
or  newly emerging flux with or without reconnection 
(e.g., Vr\v{s}nak \cite{vrsnak90}; Chen \& Shibata \cite{chen}; 
Lin et al. \cite{lin}; Zhang \& Low \cite{zhanglow})
that lead to loss of equilibrium or to ideal/resistive instabilities.  
When investigating which mechanism may be responsible for this onset, 
several authors studied the morphology, dynamics, associated brightenings, 
and their relative timing in filament eruptions in detail 
(e.g., Sterling \& Moore \cite{sterling04}, \cite{sterling05}; 
Williams et al. \cite{williams}). The results of these observational 
studies have not converged yet, and indeed they imply that a combination 
of several mechanisms may be at work in the same event. 

Oscillation of erupting filaments provides an alternative tool 
for probing the onset mechanisms. 
Filament oscillation can be classified into two groups
(Oliver \& Ballester \cite{oliver}), namely  
small-amplitude oscillations (velocity < 2-3 km s$^{-1}$) 
and large-amplitude ones (velocity $\sim $ 20 km s$^{-1}$). 
The large-amplitude oscillations are caused by 
such flare-related disturbances as Moreton waves 
(Ramsey \& Smith \cite{ramsey}; Eto et al. \cite{eto}), 
EIT waves (waves observed by the Extreme-Ultraviolet 
Imaging Telescope on SOHO, Okamoto et al. \cite{okamoto}), 
and nearby microflares (Jing et al. \cite{jing}). 
Observations of such large-amplitude oscillations can be used to 
examine the physical parameters and the equilibrium properties of 
filaments (Hyder \cite{hyder}; Kleczek \& Kuperus \cite{kleczek}; 
Vr\v{s}nak \cite{vrsnak84}). 
However, the number of reported observations of large-amplitude 
oscillations is still small. Moreover, most of them are of quiescent filaments, 
while there are few observations of the oscillation of active filaments 
(Vr\v{s}nak et al. \cite{vrsnak}). 

In this letter we report the observation of a large-amplitude oscillation 
and the subsequent eruption of a polar crown filament. 
The oscillation occurred after the filament started to rise slowly, 
and sudden acceleration and eruption followed the oscillation. 
We examine the  motion and amplitude of the oscillation and discuss 
what it indicates about the onset mechanism of the eruption.


%

\section{Observation}
The oscillation and subsequent eruption of a filament was found while the 
authors were searching for the filament eruption events in the list of 
EUV post-eruptive arcades published by Tripathi et al. (\cite{tripathib}). 
The filament is a large polar crown that erupted on 15 Oct. 2002. 
A CME was observed following the eruption, but no flare was recognized in 
the GOES light curve.   
In this study we mainly use the 195 \AA \, images from the Extreme-Ultraviolet 
Imaging Telescope (EIT; Delaboudini\`{e}re et al. \cite{delaboudiniere}) 
on the {\it Solar and Heliospheric Observatory} (SOHO). 
The time cadence of 195 \AA \, images is about 12 minutes,  
and the spatial resolution is 2.6 arcsec.

The filament was also observed by the Flare Monitoring 
Telescope (FMT; Kurokawa et al. 1995) at Hida Observatory, 
Kyoto University. The FMT observes the full disk of the Sun 
in 5 channels including H$\alpha$ line centre and wings at $\pm 0.8$ \AA \, 
with a time cadence of 2 s and a spatial resolution 
of 4.25 arcsec. Unfortunately, due to frequent interruption by clouds, 
the oscillatory motion is not clear, and the eruption occurred 
during the night at Hida observatory. However, the H$\alpha \pm 
0.8$ \AA \, images provide useful information on the line-of-sight 
motion of the filament. 

\section{Results}

Figure 1 shows the EIT 195 \AA\ images of the filament, 
in its pre-eruption and eruption phases. 
The filament has an arch-like shape in the pre-eruption phase, 
with the ends near (-100,-500) and (-700,-600) in the solar disk 
coordinate system. 
The oscillatory motion was seen only in a part of 
the filament indicated by the box in Fig. 1.

\begin{figure}
\resizebox{\hsize}{!}{\includegraphics{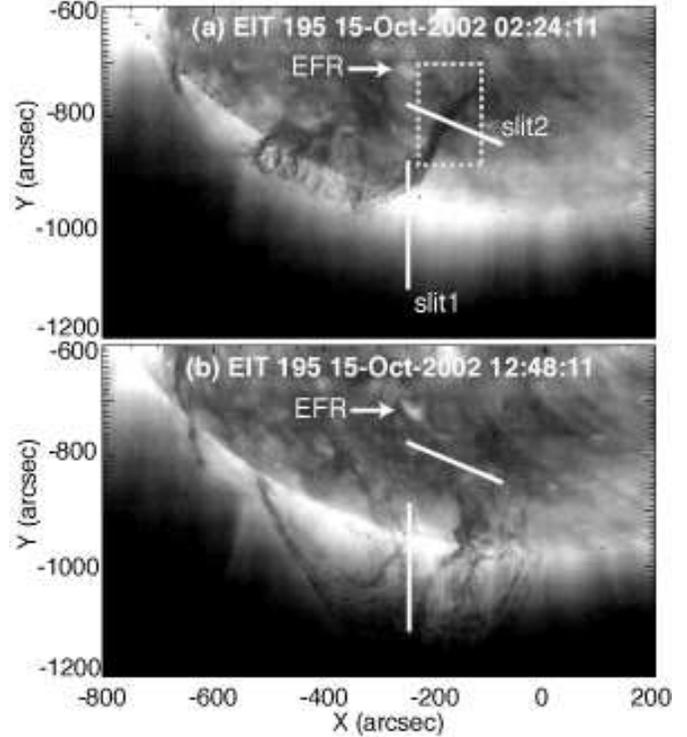}}
\caption{EIT images of the erupting filament:  
(a) pre-eruption phase and (b) eruption phase. 
The solid lines indicate the slit positions of time slices 
shown in Fig. 2. The dashed-line box shows the field of view of Fig. 3. 
The arrow points to the emerging flux region (EFR) near the filament.}
\label{fig1}
\end{figure}

To illustrate the motion of filament, Fig. 2 shows the 
intensity evolution along the slits 1 and 2 in Fig. 1 (time slices), 
corrected for solar rotation. 
The time slice for slit 1 shows that the filament 
was slowly rising for more than 10 hours until it 
was suddenly accelerated at around 11:30 UT followed by the 
eruption. The rising velocity of this pre-eruption phase is 
about 1 km s$^{-1}$. The oscillatory motion is not evident  
at the position of slit 1. 
On the other hand, the time slice for slit 2 clearly shows 
the oscillation, as well as the general trend 
to slow rising. 
The oscillation started at about 02:50 UT 
and was repeated three times before the filament finally erupted. 
The existence of the fourth oscillation is not clear because the eruption 
(strong acceleration) started after the third oscillation. 
As indicated in Fig. 2, the velocity of the oscillation was 
about 4-5 km s$^{-1}$, and 
the amplitude of displacement was about 20-30 arcsecs (not 
indicated in the figure). Note that the velocities and 
displacements are measured on the plane of the sky. 
\begin{figure}
\resizebox{\hsize}{!}{\includegraphics{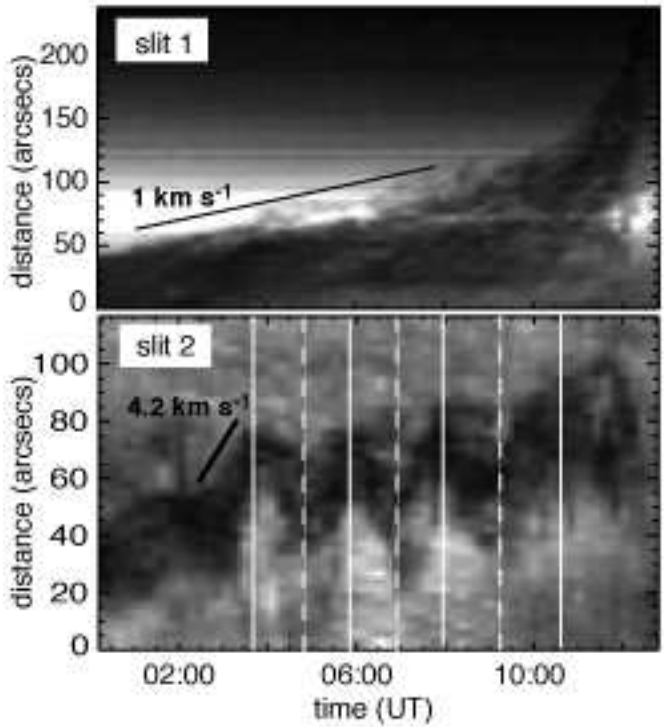}}
\caption{Time slices for slit 1 (upper panel) and slit 2 (lower panel). 
The positions of the slits are indicated in Fig. 1.
The distance is measured from the northern end of each slit. 
The black solid lines show the velocity of the filament along the slits. 
The vertical white lines in the lower panel indicate the reversal  
points of the oscillation; the solid and dashed lines correspond 
to the opposite phases.}
\label{fig2}
\end{figure}

We did not find any flares or microflares at the onset of 
the oscillation.
It was possibly triggered by the emerging 
flux under the oscillating part of the filament, which 
is indicated in Fig.1. 
The emerging flux was first identified as a growing bipolar region in 
magnetograms from the Michelson Doppler Imager aboard SOHO 
at 19:15 UT on 14 Oct. 2002. 
As a coronal counterpart, this growing bipole was seen as bright loops 
in the EIT images. 
Emerging fluxes often undergo reconnection with a pre-existing coronal field 
to produce small flares and jets (e.g., Isobe et al. \cite{isobe}). 
However, the intensity of the emerging flux in EIT images did 
not increase for several hours before and after the 
onset of oscillation. Probably the reconnection did occur and 
excited the oscillation, 
but the plasma heating and/or chromospheric 
evaporation were too weak to increase the coronal EUV emission, 
due to the weak magnetic field in this region 
(Yamamoto et al. \cite{yamamoto}; Vr\v{s}nak \& Skender \cite{vrsnak05}).

The vertical white lines in Fig. 2 indicate the times at 
which the direction of the oscillating motion seems to reverse. 
The reversal points are not clear from the time slices, particularly 
for the fourth oscillation where the eruption started. 
If we assume that the white lines in Fig.2 give the correct 
reversal points, the period of the oscillation increases slightly 
from 2 hours to 2.6 hours. This indicates the weakening of the  
restoring force, which may also be related to the destabilization 
and onset of the filament eruption. 
However, considering the uncertainty of the reversal points  
due to the internal structure of the filament and the course 
cadence (12 min) of EIT data, 
we hesitate to make any conclusive suggestions about the 
period increase based only on these data.

\begin{figure}
\centering
\resizebox{\hsize}{!}{\includegraphics{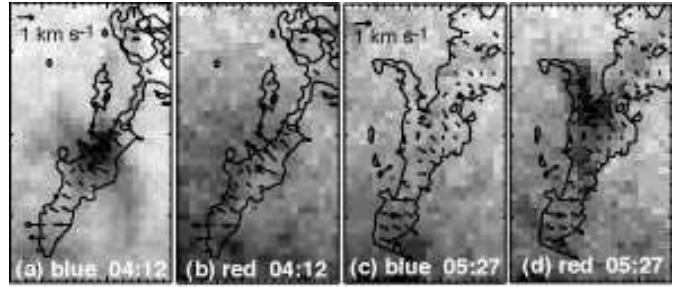}}
\caption{H$\alpha$ wing images at 04:12 UT (panels a and b) and 
at 05:27 UT (panels c and d). Panels a and c are of the blue wing 
at -0.8 \AA \, and panels b and d are of the red wing at +0.8 \AA.
Overlaid are 40 DN s$^{-1}$ pix$^{-1}$ contour
at 04:12 UT (panels a and b) and at 05:24 UT (panels c and d). 
The arrows 
show the velocity field on the plane of the sky calculated by 
correlation tracking of the EIT images. The velocity field is 
shown only in the region where the EIT count is less than 
40 DN s$^{-1}$ pix$^{-1}$.}
\label{fig3}
\end{figure}

The three-dimensional motion of the filament can be probed 
by combining the motions in the images and the information 
of Doppler shift. Figure 3 shows the images of the blue wing 
(-0.8\AA; panels a and c) and the red wing 
(+0.8\AA; panels b and d) 
of H$\alpha$ taken by FMT at 04:12 UT and 05:27 UT.  
The field of view is the same as the box in Fig. 1. 
The overlaid contours show the EIT intensity extracted from the 
EIT images recorded at the nearest times. 
The contour level is 
40 DN s$^{-1}$ pix$^{-1}$, tracing the darkest part of 
the filament where the oscillation is evident.\footnote{
see Delaboudini\`{e}re et al. (\cite{delaboudiniere}) for the definition 
of DN.}
The arrows show the velocity field on the plain of the sky 
calculated by correlation tracking.
The velocity field shows that the oscillatory motion 
is predominantly perpendicular to the filament axis. 
Also, the phase of the oscillation is nearly the 
same along the filament, i.e., the oscillating 
part of the filament moves like a rigid body. 
The other parts of the filament, those outside the box 
in Fig. 1, do not exhibit any oscillatory motion.

At 04:12 UT (panels a and b) the filament is seen as a dark feature 
in the blue wing, whereas it is almost invisible in the red wing. 
Thus it is moving towards the observer. The motion on the plane of sky 
is northeastward at this time. At 05:27 UT (panels c and d) 
when the motion on the plane  
of sky becomes southwestward, the filament appears in the red wing 
and disappears in the blue wing. 
This is the well-known feature of the ``winking filament'' 
(Hyder \cite{hyder}; Okamoto et al. \cite{okamoto}). 

Morimoto \& Kurokawa (\cite{morimoto}) have developed a method to determine the 
line-of-sight velocity from the contrasts of H$\alpha$ wing images, 
and applied the method to 5 filament eruptions observed by the FMT. 
In this event the intensity of the darkest part in blue wing image is 
about 35 \% less than the surrounding region at 04:12 UT. If we assume that 
the physical parameters of the filament such as the source function, 
optical depth, and line width are similar to those of the filaments 
analysed by Morimoto \& Kurokawa (\cite{morimoto}), the contrast of 
35 \% in the FMT blue wing corresponds to  
the line-of-sight velocity of 20-30 km s$^{-1}$, which is 
several times larger than the velocity on the plane of sky. 

Since the filament was located near the southern limb, 
the above results indicate that the oscillatory motion is 
predominantly horizontal (Kleczek \& Kuperus \cite{kleczek}). 
If the motion is purely vertical, the blue and red shifts of 
H$\alpha$ should be accompanied by a southward 
and northward motion, respectively, whereas the data show 
the blue shift with the northeastward motion and the red shift 
with the southwestward motion. This is consistent
with the predominantly horizontal motion. 

According to the model by Kleczek \& Kuperus (\cite{kleczek}), 
the period $P$ of the horizontal oscillation of a filament is given by 
$P=4 \pi L B^{-1} \sqrt{\pi \rho}$, 
where $2L$ is the length of the oscillating filament, 
$\rho$ the mass density, and $B$ 
the strength of the effective magnetic field that causes  
the restoring force of the oscillation; namely the period 
is of the order of the Alfv\'{e}n transit time of the filament. 
Assuming $\rho=10^{-13}$ g cm$^{-1}$, $B$ and 
the Alfv\'en velocity $V_A=B / \sqrt{4\pi \rho}$
can be calculated from the measured values of 
$L=10^{10}$ cm and $P=7200$ s; 
$B=1.4$ Gauss and $V_A =12$ km s$^{-1}$.  
These are reasonable values considering that the filament 
is located in the polar region where the magnetic field is weak. 

\section{Discussion}
A slow rise in the pre-eruption phase and the subsequent sudden acceleration 
and eruption are commonly observed
not only in filament eruptions but also in X-ray plasmoid ejections 
and coronal mass ejections (Ohyama \& Shibata \cite{ohyama}; 
Zhang et al. \cite{zhang}; Sterling and Moore \cite{sterling05}). 
The large amplitude of the oscillation reported in this letter 
suggests that, even during the slowly rising phase, the filament still retains 
the equilibrium, and furthermore the equilibrium is {\it nonlinearly} stable. 
Thus the slow rise in the pre-eruption phase is a 
quasi-static evolution, rather than the very slow linear stage 
of instability. 

Furthermore, the time scale of the transition from such nonlinearly stable 
equilibrium to the fast eruption by instability or loss of equilibrium 
was quite short, at most the Alfv\'{e}n transit time of the filament 
($\sim$ 1 hour). 
This suggests that the fast eruption was triggered by the fast magnetic 
reconnection that changes the equilibrium property of the 
magnetic system, rather than by the slow shearing/converging motions 
at the photosphere. However, the energy accumulation and 
slow rise may be driven by the photospheric motions. 
Such destabilizing reconnection can occur either at the 
overlying arcade above the filament (Antiochos et al. \cite{antiochos}) 
or near the footpoint (Chen \& Shibata \cite{chen}; Moore et al. \cite{moore}; 
Tripathi \cite{tripathit}).  
The emerging flux may also trigger the 
eruption withought reconnection (Vr\v{s}nak \cite{vrsnak90}; 
Lin et al. \cite{lin}). However, we did not find significant increase 
in the magnetic flux in the emerging flux region at the time of 
eruption onset, so probably 
it cannot explain the fast transion 
from oscillation to eruption observed in this event. 
We should note that in this event only a part of the 
filament exhibited the oscillation, so the equilibrium properties 
of the whole filament system may not be nonlinearly stable during 
the observed oscillation. Further search for such an oscillation of 
erupting filaments is desirable. 

The increase in the oscillation period, though not conclusive, indicates 
that the restoring force became weak during the oscillation.  
Hence, it may correspond to the destabilizing process for the eruption, and 
may be consistent with the destabilizing reconnection as the triggering 
mechanism. 
It is interesting to see whether current analytical and numerical 
models of erupting filaments can oscillate with large amplitude and, 
if so, how the filament's properties as an oscillator change 
with time before and at the onset of the eruption.

\begin{acknowledgements}

{The authors thank M. Kadota, H. Komori, and T. Okamoto 
for their help in preparing and analysing the FMT data, 
and H. E. Mason for useful comments. 
HI is supported by the Research Fellowship from the Japan 
Society for the Promotion of Science for Young Scientists. 
DT acknowledges funding from PPARC. 
SOHO is a project of international collaboration of ESA and NASA. Moreover, we 
would also like to acknowledge Castle, a pub in Cambridge,  
where the idea of this paper first came up and was developed. 
}
\end{acknowledgements}


\end{document}